\documentclass[prb,twocolumn,superscriptaddress]{revtex4}

\usepackage{graphicx}

\begin{document}
\preprint{cond-mat\000000}

\title{Measurements of the absolute value of the penetration depth in
high-$ T_c$ superconductors using a tunnel diode resonator}

\author{R. Prozorov}
\author{R. W. Giannetta}
\affiliation{Loomis Laboratory of Physics, University of Illinois at
Urbana-Champaign, 1110 West Green St., Urbana, 61801 Illinois.}

\author{A. Carrington}
\affiliation{Department of Physics and Astronomy, University of Leeds, Leeds
LS2 9JT, United Kingdom}

\author{P. Fournier}
\author{R. L. Greene}
\affiliation{Center for Superconductivity Research, Department of Physics,
University of Maryland, College Park, Maryland 20742 }

\author{P. Guptasarma}
\author{D.G. Hinks}
\affiliation{Chemistry and Materials Science Division, Argonne National
Laboratory, Argonne, Illinois 60439}

\author{A. R. Banks}
\affiliation{Materials Research Laboratory, University of Illinois at Urbana-
Champaign, 104 South Goodwin Ave, Urbana, 61801 Illinois.}

\date{Submitted to Applied. Phys. Lett. June 30, 2000}

\begin{abstract}
A method is presented to measure the absolute value of the London penetration
depth, $\lambda$, from the frequency shift of a resonator. The technique
involves coating a high-$T_c$ superconductor (HTSC) with film of low - $T_c$
material of known thickness and penetration depth. The method is applied to
measure London penetration depth in YBa$_2$Cu$_3$O$_{7-\delta}$ (YBCO)
Bi$_2$Sr$_2$CaCu$_2$O$_{8+\delta}$ (BSCCO) and
Pr$_{1.85}$Ce$_{0.15}$CuO$_{4-\delta}$ (PCCO). For YBCO and BSCCO, the values
of $\lambda (0)$ are in agreement with the literature values. For PCCO
$\lambda \approx 2790$ \AA, reported for the first time.
\end{abstract}

\pacs{74.20.De, 74.25.Ha, 74.25.Nf}

\maketitle

The London penetration depth, $\lambda(T)$, is a quantity of fundamental
importance. Its temperature, field, and doping dependencies are directly
related to the density of quasiparticle energy states, knowledge of which is
crucial for testing models of pairing symmetry and mechanisms of
superconductivity \cite{annett90}. $\lambda(T)$ is also a key parameter in
determining the response and collective properties of the Abrikosov vortex
lattice \cite{blatter94}. For single crystals, measurements of the resonant
frequency shift of a microwave cavity
\cite{hardy93,anlage94,hosseini98,kamal98} or tunnel diode oscillator
\cite{carrington99a,carrington99,prozorov00,prozorov00a} provide the highest
resolution for changes of the penetration depth, $\Delta \lambda \equiv
\lambda(T) - \lambda(T_{min})$, with respect to temperature. For the sub-mm
sized crystals typically studied in high-$T_c$ work, resolution of better than
$0.2$ \AA\ can be achieved \cite{hardy93,kamal98,carrington99,prozorov00}.
However, the usual resonator approach has the disadvantage that it does not
provide the absolute magnitude of $\lambda$. This shortcoming arises from
various experimental uncertainties and is not an inherent limitation of the
resonator technique. As we show in this paper, by suitably plating
superconducting crystals it is possible to exploit the extremely high
sensitivity of the resonator to changes in frequency and thus obtain an
absolute measurement of $\lambda(T)$.

The method described here permits a simultaneous measurement of $\lambda
(T_{min})$ and $\Delta \lambda(T)$ on the same sample. The zero-temperature
penetration depth, $\lambda(0)$ can be obtained by extrapolation to $T=0$.
Together, $\lambda (0)$ and $\Delta \lambda (T)$ determine the normalized
superfluid density $\rho_s(T)=(1+\Delta \lambda (T)/\lambda (0))^{-2}$, the
quantity directly related to the electromagnetic response of the
superconductor. This is a distinct advantage over the situation in which these
two quantities are obtained by different groups using different samples and
techniques. In addition, no new physical model is required to obtain
$\lambda(T)$ from the data, unlike the case with techniques such as $\mu$SR or
reversible magnetization. Our method has been tested on single crystals of
YBCO, BSCCO and PCCO and compared with values of $\lambda(T)$ obtained from
other techniques.

It is first worth discussing why resonator methods cannot normally determine
the absolute penetration depth. We focus on a lumped LC resonator but the
ideas also hold for a distributed device such as a microwave cavity. In the
absence of a superconducting sample the empty resonant frequency is $f_0 =
1/\sqrt{LC}$. When a superconducting sample is inserted into the resonator,
the inductance $L$ decreases due to a decrease of the magnetic field energy
$W_m =LI^2/2c^2$ as a result of Meissner expulsion. For a platelet sample of
thickness $2d$ in the $z-$ direction and mean planar dimensions $2w \times 2w$
in the $x-y$ plane, this leads to an increase of the frequency by an amount $
\Delta f \equiv f(T)-f(0) $ given by \cite{prozorov00},

\begin{equation}
\label{df} \frac{\Delta f}{f_{0}}=\frac{V_s}{2V_0 \left( 1 - N\right) }\left[
1-\frac{\lambda }{R}\tanh \frac{R}{\lambda }\right]
\end{equation}

Here $V_s$ is the sample volume, $V_0$ is the effective volume of the
resonator, $N$ is the demagnetization factor and the field is applied along
the $z$ direction. $R$ is the effective sample dimension which depends upon
field orientation relative to the sample and sample geometry \cite{prozorov00}.
For the standard ``Meissner`` configurations in which the field is applied
parallel to the surface of an infinite slab, $N = 0$ and $R = d$. For the
geometry used here, in which the AC field is normal to the face of a platelet,
$R \approx w/5$ \cite{prozorov00}.

\begin{figure} [tb]
\includegraphics[width=8cm,keepaspectratio=true]{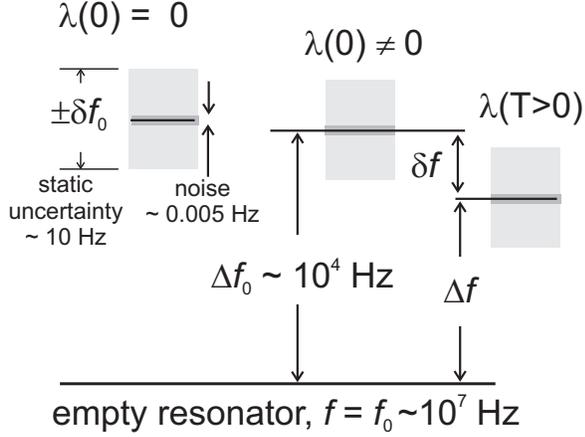}
 \caption{Frequency shifts encountered in resonance measurements. The static uncertainty
 $\delta f_0$ is usually greater than the frequency shift due to finite $\lambda (0)$. Relative
 frequency shift $\delta f$ is independent of $\delta f_0$ and permits accurate measurements
 of $\Delta \lambda$.}
 \label{scheme_of_df}
\end{figure}

The measurement process is sketched in Fig.~\ref{scheme_of_df}. The
superconducting sample is inserted into the resonator, resulting in a change
in frequency $\Delta f_{0}$. For typical HTSC samples $ \Delta f_{0} ~ 10^4$
Hz. In principle, if $R$ were known precisely then one could use
Eq.~(\ref{df}) together with the measured $\Delta f_{0}$ to determine
$\lambda(0)$. Unfortunately, there are several sources of error. First, the
accuracy with which $ \Delta f_{0}$ can be determined is limited by
repeatability. Extracting and inserting the sample in situ typically leads to
an error of $\delta f_0\sim 10$ Hz out of a total $\Delta f \sim 10^4$. This
``static`` uncertainty is shown by the gray band in Fig.~\ref{scheme_of_df}.
 According to Eq.~(\ref{df}), the difference between the perfect
diamagnet and sample with finite $\lambda$ is only $f_0 (1-\lambda/R) \approx
30$ Hz for a typical YBCO crystal where $R \geq 50$ $\mu$m and $\lambda(0) =
0.15$ $\mu$m which is quite comparable to the static uncertainty, $\delta f_0
\approx 10$ Hz. Furthermore, extracting and inserting the sample gives the
value of $\Delta f_0$ already reduced by finite $\lambda (0)$. Other methods of
estimation of $\Delta f_0$ such as measuring a ball made of a conventional
superconductor \cite{carrington99a,carrington99} or replicating an HTSC sample
using low-$T_{c} $ material \cite{hardy93} result even in greater uncertainty,
because in addition to an inevitable ``static`` uncertainty between different
runs there are additional uncertainties related to differences between real
sample and the substitute. Furthermore, realistic samples are irregular and so
have dimensions which are uncertain to much more than $\lambda(T)$. They may
also have large demagnetizing effects. Finally Eq.~(\ref{df}) itself involves
approximations for $R$ that adds further error. It is therefore not feasible
to measure $\lambda(T)$ using resonator frequency shifts in the
straightforward manner outlined. Despite this limitation on accuracy, the
precision with which changes in $\lambda$ can be measured is much higher. This
is illustrated in Fig.~\ref{scheme_of_df} by the change $\Delta f$ upon
warming the sample from low to intermediate temperature. In this case the
sample stays fixed so the temperature-independent static uncertainty is
irrelevant. Only the oscillator noise matters, which is typically 2000 times
smaller than the static uncertainty. It is therefore imperative to adopt a
technique that keeps the sample fixed.

\begin{figure} [tb]
\includegraphics[width=8cm,keepaspectratio=true]{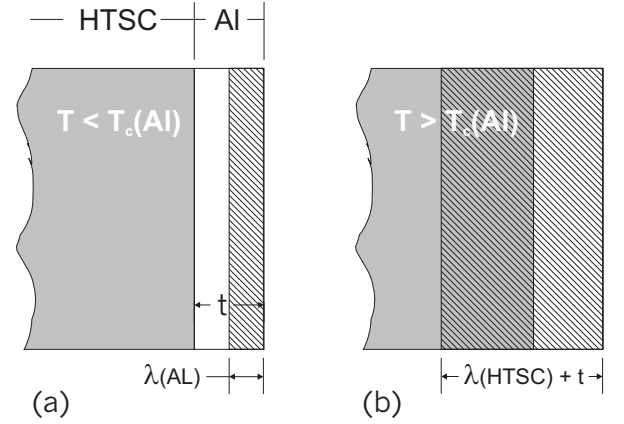}
 \caption{Field penetration above and below transition temperature of Al film. (a) $T < T_c(Al)$: magnetic field penetrates only $\lambda (Al)$.
 (b) $T > T_c(Al)$: magnetic field penetrates the Al layer and HTSC superconductor $\lambda (HTSC)$.}
 \label{scheme}
\end{figure}

Our method is illustrated in Fig.~\ref{scheme}. The sample under study is
plated with a conventional low $T_c$ superconductor, in this case an Al film.
The film thickness $t$ should be larger than $\lambda(Al)$ but much smaller
than the normal state skin depth of Al ($ \approx 3$ $\mu$m at the operating
frequency of 10 MHz).

Using $\lambda = H^{-1} \int_0^{\infty}{B(x)dx}$ we find that for $T < T_c(Al)$
the magnetic field penetrates to $\lambda (T<T_c(Al))= \lambda(Al)+
\exp{(-t/\lambda(Al))} \left[ \lambda(HTSC) -\lambda(Al) \right]$. Above
$T_c(Al)$ the penetration depth is $\lambda (T>T_c(Al))= t+\lambda(HTSC)$.
Converting the frequency change $\Delta f = f(T>T_c(Al))-f(T<T_c(Al))$ to a
change in the effective penetration depth, $\Delta \lambda$, using
Eq.~(\ref{df}), we obtain:

\begin{equation}
\lambda (HTSC) = \lambda (Al) + \frac{\Delta \lambda - t}{1 - \exp{(-
t/\lambda (Al))}} \label{dl}
\end{equation}

The errors in this method arise from uncertainties in the film thickness $t$,
the resonator calibration constant and $\lambda (Al)$. Literature values for
the effective penetration depth of aluminum films, $\lambda (Al) \approx
\lambda_L(Al) \sqrt{\xi (0)/\ell}$ range from $400$ to $600$ \AA
\cite{aluminumlambda}. Here, the BCS coherence length $\xi (0) \approx 16000$
\AA, the mean free path, $\ell \approx 1000$ \AA\ and the London penetration
depth $\lambda_L(Al) \approx 160 $ \AA. We choose the commonly accepted value,
$\lambda (Al) \approx 500 \pm 100 $ \AA. The Al film was $800 \pm 50 $ \AA\
thick. Uncertainty in the calibration constant gives an additional error of
about $10 $ \AA\ giving a total error of approximately $\pm 150 $ \AA. It can
be further reduced by choosing different coating materials, which will give
additional independent reference points and by varying the thickness of the
coating layer. Although it is clearly desirable to improve the accuracy, an
error of $150 $ \AA\ still results in only a $1 \%$ deviation of $\rho_s$ over
20 K range for YBCO. A somewhat similar measurement technique was used earlier
to determine $\lambda$ in heavy fermion compounds \cite{gross89,grob91}. In
that experiment, a flux trapped or screened by a thin Cd layer was used, but
owing to the much reduced sensitivity of SQUID magnetization measurements, it
is not suitable for high-$T_c$ materials.

$\lambda (HTSC)$, was measured in three different superconductors:
YBa$_2$Cu$_3$O$_{7-\delta}$ (YBCO), Bi$_2$Sr$_2$CaCu$_2$O$_{8+\delta}$
(BSCCO), and the electron-doped Pr$_{1.85}$Ce$_{0.15}$CuO$_{4-\delta}$ (PCCO).
YBCO crystals were grown in yttria stabilized zirconia crucibles as described
\cite{ybco} and annealed to achieve maximal $T_c \approx 93$ K \cite{ybco}.
BSCCO samples where grown using a floating zone process and had $T_c \approx
89.5 K $\cite{pcco}. Single crystals of PCCO were grown using directional
solidification technique and annealed in argon to achieve $T_c \approx 22.5 K$
\cite{pcco}. The aluminum coating was applied with a magnetron sputtering
system with 5 cm rotated Al target ( 99.999 \% purity). Sputtering was
conducted in an argon atmosphere and was homogeneous over 20 cm$^2$. The Al
layer thickness, $t$, was calibrated using a {\it Inficon XTC 2} with 6 MHZ
gold quartz crystal and later directly measured using SEM edge imaging of a
broken sample.

The measurement technique utilized a 10 MHz tunnel diode oscillator whose
specifications have been reported previously \cite{carrington99a,prozorov00}.
Samples were mounted on a moveable sapphire stage whose temperature could be
varied from 0.35 to 100 K. The low base temperature was crucial in order to
obtain the full frequency shift due to the diamagnetism of the Al film.

\begin{figure} [tb]
\includegraphics[width=8cm,keepaspectratio=true]{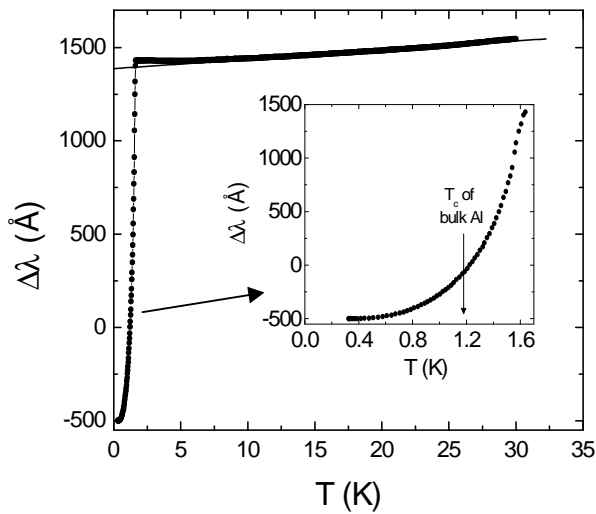}
 \caption{Penetration depth in single crystal YBCO calculated from Eq.~(\ref{dl}).
 \textit{Inset:} Temperature range in which Al becomes normal.}
 \label{ybco}
\end{figure}

\begin{figure} [tb]
\includegraphics[width=8cm,keepaspectratio=true]{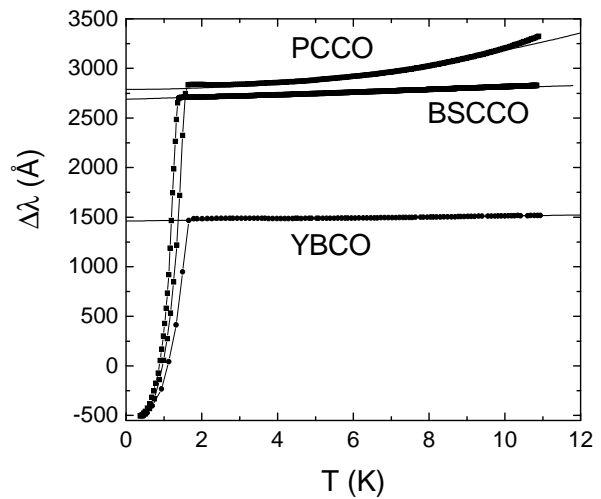}
 \caption{Penetration depth in (from bottom to top) YBCO, BSCCO and PCCO single crystals.
 For YBCO and BSCCO data were extrapolated using $\lambda(T) \propto T$, whereas for PCCO
 $\lambda (T) \propto T^2$ dependence was used.}
 \label{data}
\end{figure}

We first present experiments in YBCO single crystals. Previous work has shown
that $\lambda(0)$ is anisotropic with $\lambda_a(0)= 1600$ \AA\ and
$\lambda_b(0)= 800$ \AA \cite{kamal98}. Since supercurrents for the $H || c$
orientation flow along both $a$ and $b$ axes, we obtain an average of
$\lambda_a$ and $\lambda_b$. Two crystals, plated in separate evaporation runs,
were measured. The first is shown in Fig.~\ref{ybco}. Note that contribution
due to Al film is subtracted using Eq.~(\ref{dl}) and therefore the $\Delta
\lambda (T)$ curve begins at negative values. Thus, at $T=T_c(Al)$, $\lambda
(HTSC)(T_c(Al))$ is obtained. Linear extrapolation to $T=0$ yields $\lambda
(YBCO) \approx 1460$ \AA. This value should be compared to values obtained
from $\mu$SR, $1405 \pm 92$ \AA\ \cite{bonn}, $1550$ \AA\
 \cite{tallon}, $1586-1699$ \AA\ \cite{sonier}; polarized neutron
reflectometry, $1400 \pm 100 $ \AA\ \cite{lauter}; magnetic susceptibility of
grain - aligned powder, $1400$ \AA\ \cite{panagopoulos} and infrared
spectroscopy, $1440$ \AA\ \cite{basov}. Since $T_c(Al)$ is quite low and the
Al plating in its normal state is transparent to 10 MHz RF, it is possible to
monitor $d\lambda (T)/dT$ of YBCO for $T > T_c (Al)$. This is an important
check on the method since it is conceivable that the Al coating might change
the surface properties of the cuprate enough to alter its penetration depth.
The slope $d\lambda /dT \approx 5.1$ \AA/K. This slope is somewhat larger than
the value of $4.1$ \AA/K reported previously
\cite{hardy93,kamal98,carrington99a}, but is in agreement with our recent
measurements conducted on unplated samples in the $H || ab$ configuration. The
second YBCO sample, shown in Fig.~\ref{data}, gave $\lambda (0) \approx 1460$
\AA and $d\lambda /dT \approx 5.10$ \AA/K, both within the estimated error
with the first sample.

The Inset to Fig.~\ref{ybco} shows details of the penetration depth variation
warming the sample above $T_c(Al)$. The measured $T_c(Al) \approx 1.69$ K is
significantly larger than the bulk value $T_c(Al) \approx 1.18$ K
\cite{aluminumtc}. This increase could be due to proximity effects
\cite{farrell89}, but could also be caused by disorder and altered chemical
composition of aluminum film \cite{aluminumtc}.

Figure \ref{data} summarizes the measurements for all three cuprates. For
BSCCO-2212 crystal we obtained $\lambda(\text{BSCCO}) \approx 2690$ \AA, which
can be compared to data from: reversible magnetization, $\lambda \approx 2100$
\AA\ \cite{kogan}; $\mu$SR, $\lambda \approx 1800$ \AA\ \cite{lee}; and lower
critical field measurements, $\lambda \approx 2700$ \AA \cite{niderost}. It is
clear that a fairly large disagreement still exists over the value of
$\lambda(0)$ in this material. We obtained a linear variation of $\lambda(T)$
with a $d\lambda /dT\approx 11.7$ \AA/K, compared to $d\lambda /dT\approx
10.5$ \AA/K in previous microwave and $\mu$SR \cite{jacobs,lee2}. To within
our current precision, it appears that the Al plating has no effect on the
electrodynamics of the underlying cuprate superconductor.

The uppermost curve in Fig.~\ref{data} shows the results for the
electron-doped cuprate superconductor, PCCO. This material has been cited as an
example of a cuprate s-wave superconductor. Recent measurements with higher
resolution and lower temperatures have shown that $\lambda(T)$ varies
quadratically with temperature, indicative of a nodal order parameter in the
presence of impurity scattering \cite{prozorov00a}. This is shown in the
figure with $d\lambda /dT \approx 4.38$ \AA/K$^2$. We find $\lambda (0) \approx
2790$ \AA. The only published value was obtained from measurements of
$H_{c1}=\Phi_0/[4\pi \lambda (0)^2] \ln{\kappa}$ which gave $\lambda
(0)\approx 1000$ \AA\ \cite{hoekstra}. It is difficult to reliably determine
$H_{c1}$ in thin crystals owing to demagnetization and pinning surface barrier
effects. Our approach is arguably more reliable since no DC fields or vortices
are involved and we have obtained close agreement with other methods in YBCO
and BSCCO.

In conclusion, we have developed a new technique to measure $\lambda(T)$ in
high$-T_c$ superconductors. We obtained $\lambda (0) = 1390$ and $1460$ \AA\
for two YBCO crystals, $\lambda (0) = 2690$ \AA\ for BSCCO, and $\lambda (0) =
2790$ \AA\ for PCCO. All values are determined with the $\pm 150$ \AA
accuracy. The plating has no discernable effect on the underlying temperature
dependence of $\lambda(T)$. The accuracy of the method is limited principally
by uncertainties in the Al film thickness.

\begin{acknowledgments}
We thank P. J. Hirschfeld for pointing out Ref.[\onlinecite{gross89,grob91}].
Work at Urbana was supported by the Science and Technology Center for
Superconductivity Grant No. NSF-DMR 91-20000. Work at Maryland was supported
by NSF Grant No. NSF-DMR 97-32736. Work at Argonne was supported by U.S.
DOE-BES Contract No. W-31-109- ENG-38 and NSF-STCS Contract No. DMR
91-20000\end{acknowledgments}

\end{document}